\documentclass[iop]{emulateapj}
\usepackage{graphicx,amsmath}
\usepackage{color}
\usepackage[plainpages=false, colorlinks=true, anchorcolor=blue, linkcolor=blue, citecolor=blue, bookmarks=false]{hyperref}

\newcommand{\kms}{{\rm\,km\,s^{-1}}}
\newcommand{\cs}{{c_s}}
\newcommand{\pc}{{\rm\,pc}}
\newcommand{\kpc}{{\rm\,kpc}}
\newcommand{\yr}{{\rm\,yr}}
\newcommand{\Gyr}{{\rm\,Gyr}}
\newcommand{\Msun}{{\rm\,M_\odot}}
\newcommand{\freq}{{\kms\kpc^{-1}}}
\newcommand{\surf}{{\Msun\pc^{-2}}}

\begin{document}

\title{Nuclear Spiral Shocks and Induced Gas Inflows in Weak Oval Potentials}

\slugcomment{Accepted for the publication in the ApJL}
\shorttitle{Nuclear Spiral Shocks}
\shortauthors{Kim \& Elmegreen}

\author{Woong-Tae Kim}
\affil{Department of Physics \& Astronomy, Seoul National University, Seoul 151-742,
    Republic of Korea} \email{wkim@astro.snu.ac.kr}

\author{Bruce G. Elmegreen}
\affil{IBM T. J. Watson Research Center, 1101 Kitchawan Road, Yorktown Heights, New York 10598 USA} \email{bge@us.ibm.com}

\begin{abstract}
Nuclear spirals are ubiquitous in galaxy centers. They exist not only in strong barred galaxies but also in galaxies without noticeable bars. We use high-resolution hydrodynamic simulations to study the properties of nuclear gas spirals driven by weak bar-like and oval potentials. The amplitude of the spirals increases toward the center by a geometric effect, readily developing into shocks at small radii even for very weak potentials. The shape of the spirals and shocks depends rather sensitively on the background shear. When shear is low, the nuclear spirals are loosely wound and the shocks are almost straight, resulting in large mass inflows toward the center.  When shear is high, on the other hand, the spirals are tightly wound and the shocks are oblique, forming a circumnuclear disk through which gas flows inward at a relatively lower rate. The induced mass inflow rates are enough to power black hole accretion in various types of Seyfert galaxies as well as to drive supersonic turbulence at small radii.
\end{abstract}

\section{Introduction}

\begin{figure}
\centering\includegraphics[width=3.4in]{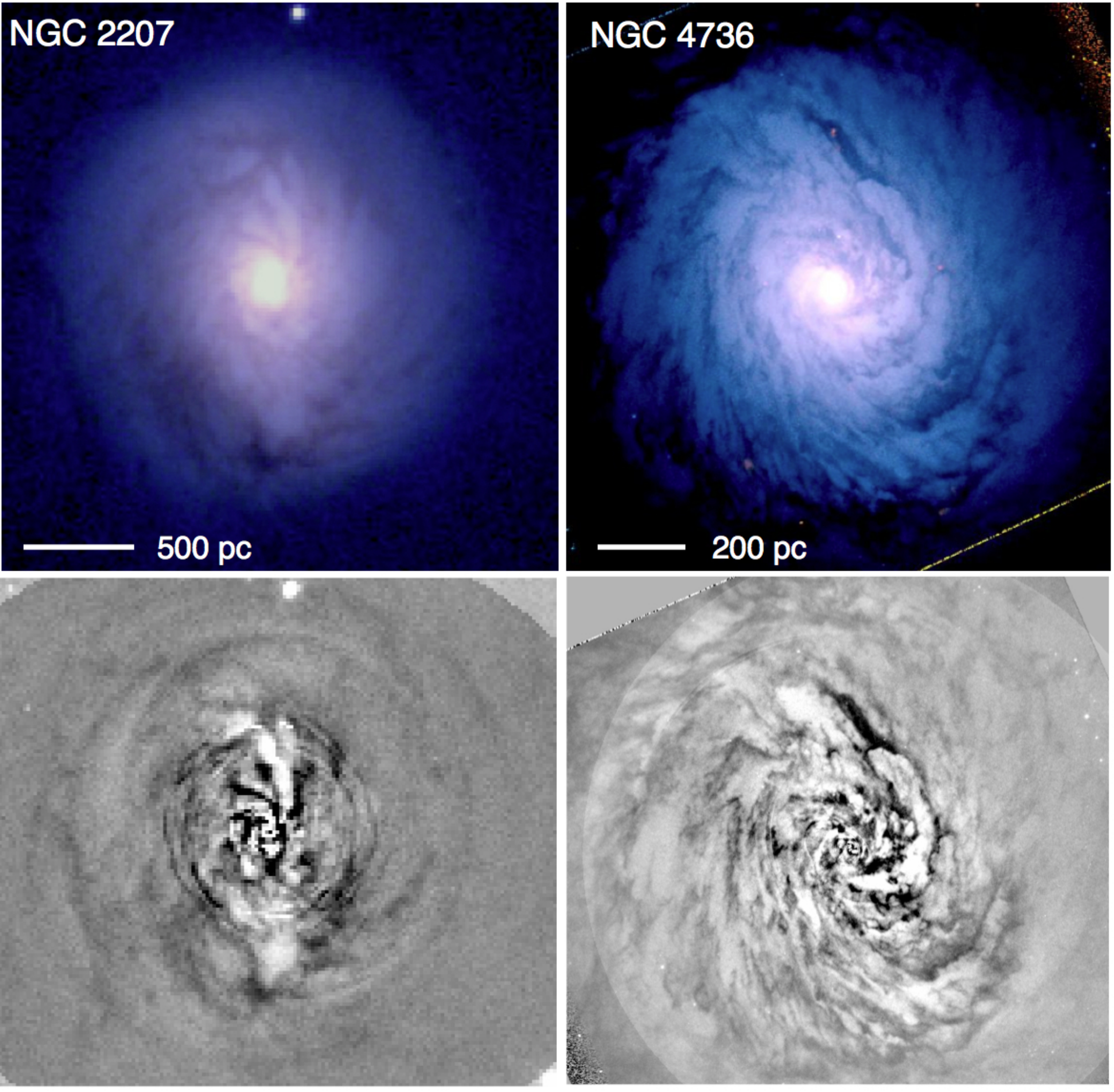}
\caption{Central regions of NGC 2207 \citep{elmegreen98} and NGC 4736 \citep{elmegreen02} observed by the \emph{Hubble Space Telescope}. The lower panels display the residuals after subtracting the ellipse fit averages from the total images shown in the top panels. Spirals in NGC 2207 are relatively loosely wound and branch out into V shape at some radii, while those in NGC 4736 are relatively tightly wound. Both galaxies have a weak bar at their centers.}\label{f:image}
\end{figure}

Nuclear dust and gas spirals are observed on scales of several hundred parsecs in the central parts of the Milky Way \citep{roberts93,sofue95,sawada04,hsieh15,ridley17} and other disk galaxies
\citep{buta93,knapen95}. Sometimes these spirals have irregular shapes
\citep{elmegreen98,martini03a,martini03b} reminiscent of turbulence
\citep{elmegreen02}. Figure \ref{f:image} shows two examples of irregular spirals. The more regular cases have been attributed to gas flows in bars or nuclear bars \citep{simkin80,shlosman89b,martini03b}, while the irregular spirals could be from gravitational \citep{shlosman89a}, magnetic \citep{balbus91}, or hydrodynamic \citep{montenegro99,krumholz15} instabilities. Since many nuclear disks have Toomre stability parameters much larger than unity \citep{elmegreen98,martini99,lin16}, the spirals in these cases are unlikely to form by two-dimensional (2D) gravitational processes.  This leaves the question of how nuclear turbulence might be generated and how it can become nonlinear to the point where shocks form, giving the appearance of sharply delineated, irregular, dust lanes.

Here we consider the possibility that turbulent wavelets moving inward will strengthen by geometric convergence, as predicted by the $1/R$ terms for radius $R$ in the nonlinear equations for spiral density waves \citep{bertin89}. \cite{montenegro99} called this a curvature instability although it is more of an amplification of pre-existing waves than an exponential growth from noise.  We simulate 2D nuclear spirals driven by oval distortions of various strengths and rates of shear in the background rotation curve. Even with a weak distortion, the resulting gas structures readily form shocks at small radii, as predicted by the analytic theory. These shocks then torque the gas and cause it to accrete at a fairly high rate. We suggest that in three-dimensional models with turbulent energy dissipation, the gravitational binding energy released by this accretion will sustain the turbulence.

There have been many previous studies of bar-driven nuclear spirals starting with \cite{sanders76}, \cite{roberts79}, \cite{athanassoula92}, \cite{piner95}, and others. Simulations by \cite{maciejewski04}, \cite{thakur09} and others traced bar-driven spiral shocks all the way to the vicinity of a nuclear black hole. These shocks were driven by strong bars and formed the usual long dust-lanes that are observed in barred galaxies. \cite{kim12a,kim12b} and \cite{li15} also had a strong bar to study the formation of nuclear rings.  Strong forcings like these overwhelm the geometric effects of shock steepening during inward wave propagation. Other works also did not consider our low-shear case, which brings out the curvature steepening even more as the resulting spiral pitch angles are large.

\section{Models}

%fig1

We consider bar- and oval-driven spirals in two models with different  background shear: HS and LS models. The HS model is to represent the central regions of galaxies with high shear like the Milky Way, whose mass distribution is dominated by a nuclear bulge and a supermassive black hole (SMBH) \citep{launhardt02}. The resulting rotation velocity (e.g., \citealt{krumholz15}) can be well fitted by
\begin{equation}\label{eq:mil}
 V=65+95\tanh\left(\frac{R-0.07}{0.06}\right) - 50\log R + 1.5 (\log R +3 )^3,
\end{equation}
for $\;0.01\leq R\leq 10$, where $V$ is in units of km s$^{-1}$ and $R$ in kpc. On the other hand, the LS model is designed to simulate disk galaxies with low shear, like NGC 3041 \citep{erroz16}, for which we take
\begin{equation}\label{eq:gal}
 V^2 = \frac{GM_{\rm BH}}{R} + \left(\frac{V_0 R}{R_0 + R}\right)^2,
\end{equation}
where $V_0=220\rm\;km\;s^{-1}$, $R_0=0.3\kpc$, and a black hole mass of $M_{\rm
BH}=3\times 10^6\;M_\odot$.

\begin{figure}[!t]
\centering\includegraphics[width=3.4in]{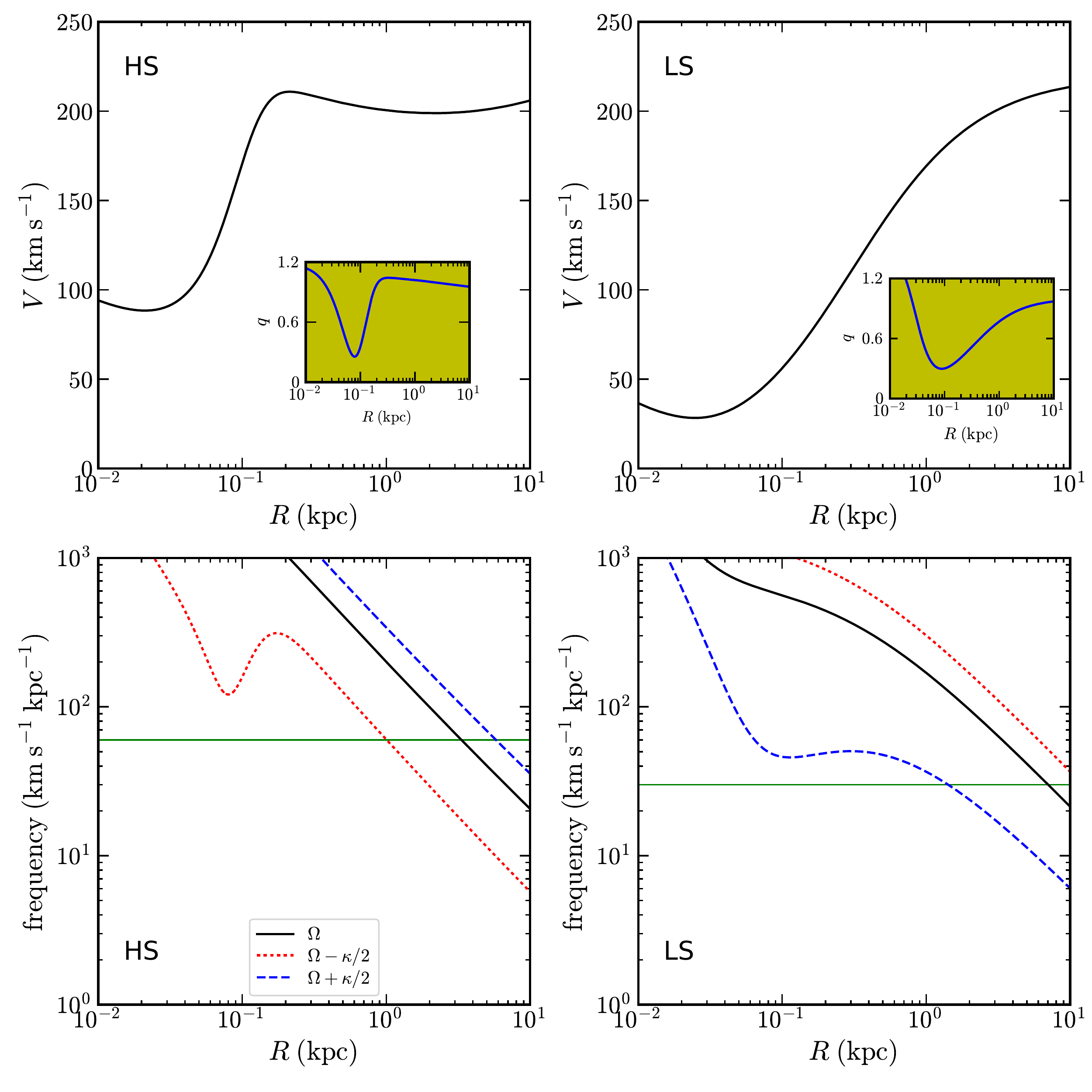}
\caption{Radial profiles of the rotation curves and various frequencies of the HS and LS models. The top panels plot the rotational velocity $V$.  The inset gives the shear parameter $q$, showing that shear is quite strong in the HS model except near $R\sim0.08$ kpc, while the LS model has relatively weak shear in the regions with $R\sim0.1-1$ kpc. The bottom panels plot $\Omega$ and $\Omega\pm\kappa/2$, where $\kappa=(4-2q)^{1/2} \Omega$ is the epicycle frequency.  The patten speed of the external potential in the HS and LS models is taken to $\Omega_p=60$ and $30\freq$, respectively, indicated as the horizontal lines.  The inner Lindblad resonance and corotation resonance are at $R_{\rm ILR}=1.00\kpc$ and $R_{\rm CO}=3.32\kpc$ in the HS model, and at $R_{\rm ILR}=1.42\kpc$ and $R_{\rm CO}=7.03\kpc$ in the LS model.}\label{f:galmodel}
\end{figure}

Figure \ref{f:galmodel} plots the radial distributions of the rotational velocity, shear parameter $q\equiv - d\ln\Omega/d\ln R$ with angular frequency $\Omega=V/R$, and various frequencies of the HS and LS models. The gas is taken to be infinitesimally thin, initially uniform with surface density $\Sigma_0$, and isothermal with a constant speed of sound $\cs$. We take $\cs=10\kms$ as our fiducial value, but also consider the cases with $\cs=20\kms$. For the perturbing gravitational potential, we adopt a simple bi-symmetric sinusoidal form  $\Phi_{\rm ptb} (R,\phi, t) = \Phi_a \cos(2\phi - 2\Omega_p t)$ with constant amplitude $\Phi_a$. We consider two cases with $\Phi_a/V_0^2=2\times 10^{-5}$ for an oval distortion in an unbarred galaxy and $\Phi_a/V_0^2=2\times 10^{-4}$ for a weak bar in a barred galaxy. These may also represent a weak secondary bar or a more-or-less isotropic central portion of a strong bar in more strongly barred galaxies. By considering a weak bar, our models cannot capture interactions of nuclear spirals with a nuclear ring that forms by a strong bar.

%fig2

The potential rotates rigidly with pattern speed $\Omega_p=60$ and $30\freq$ for the HS and LS models, respectively.  We use Athena++, a newly developed grid-based code utilizing a higher-order Godunov scheme (Stone et al. 2017, in preparation), to evolve the gas in 2D logarithmic cylindrical coordinates \citep{kim12b}. Our simulation domain extends from the inner radial boundary, $R_{\rm in}=10\pc$, to the corotation radius (CR), and covers $\phi=0-2\pi$. The number of zones in the radial and azimuthal directions is $1024\times516$ and $1024\times548$ in the HS and LS models, respectively, making all the zones approximately square-shaped.

The imposed gravitational potential perturbs an otherwise uniform gas disk. The
perturbations organize into spiral waves that grow as thermal pressure tends to align the apocenters of perturbed gas orbits inside the inner Lindblad resonance (ILR), reinforcing the perturbations \citep{montenegro99}. This is opposite to the case of spirals outside the ILR where gravity aligns the perturbed orbits and pressure resists them.

\begin{figure*}
\centering\includegraphics[width=6in]{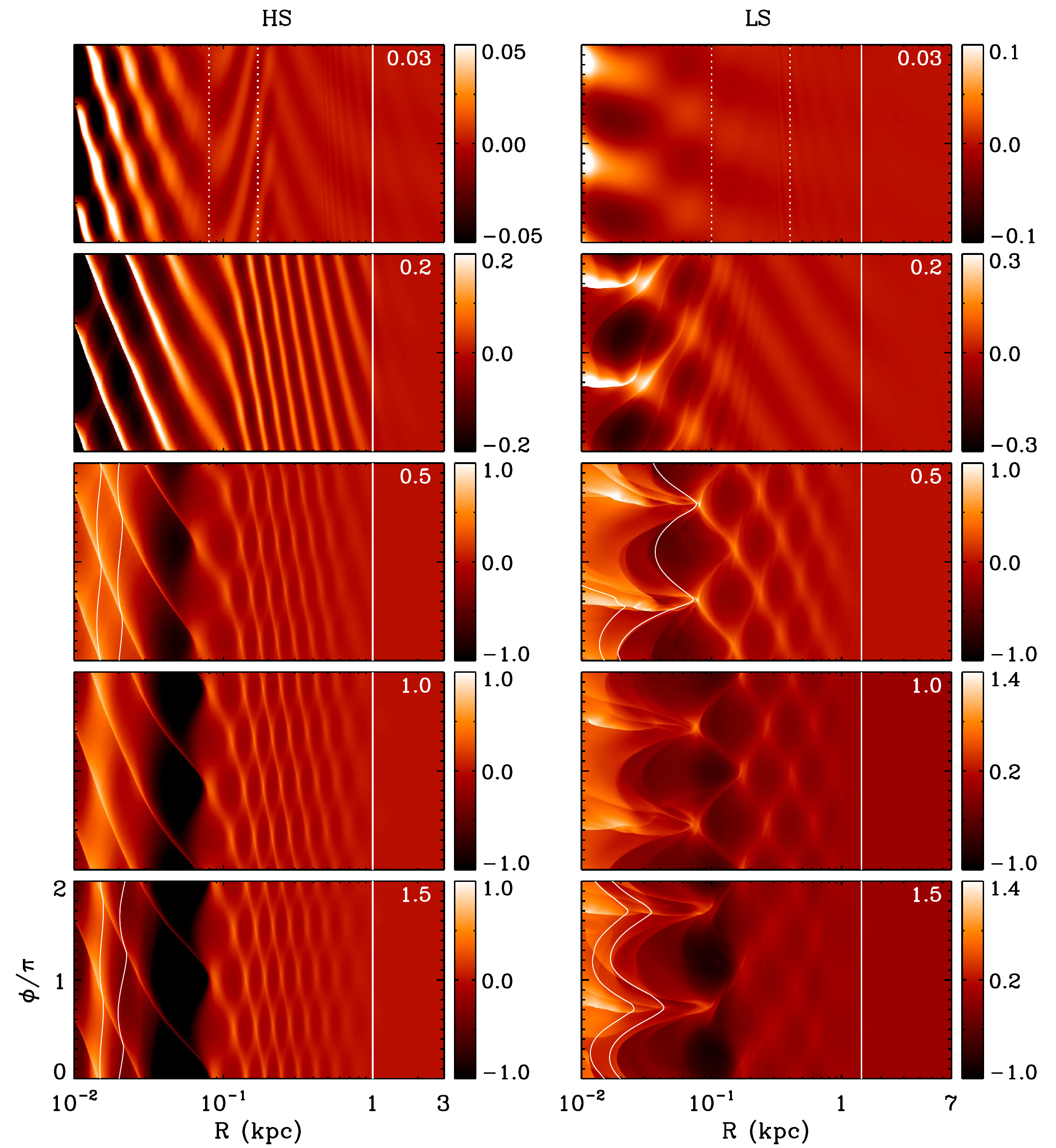}
\caption{Evolution of the gas surface density in the $\log R$--$\phi$ plane for the HS and LS models with $\Phi_a/V_0^2=2\times 10^{-4}$ and $\cs=10\kms$. The number in the upper-right corner in each panel is the time in units of Gyr. The vertical solid lines mark the ILR in all panels. In the top panels, the vertical dotted lines indicate the positions, $R_1=0.08$ kpc and $R_2=0.17$ kpc in the HS model and $R_1=0.1$ kpc and $R_2=0.4$ kpc in the LS model, where $d\Theta/dR$ changes its sign. Thin lines in the middle and bottom panels show the instantaneous streamlines running from $R=$ 15 and 20 pc at $\phi=0$ to the positive $\phi$ direction. In addition to spirals driven by the external potential, very-weak, tightly-wound spiral perturbations propagating from both radial boundaries are present in the top panels. Colorbars label
$\log(\Sigma/\Sigma_0)$.}\label{f:logRphi}
\end{figure*}

\section{Results}

\begin{figure*}
\centering\includegraphics[width=7in]{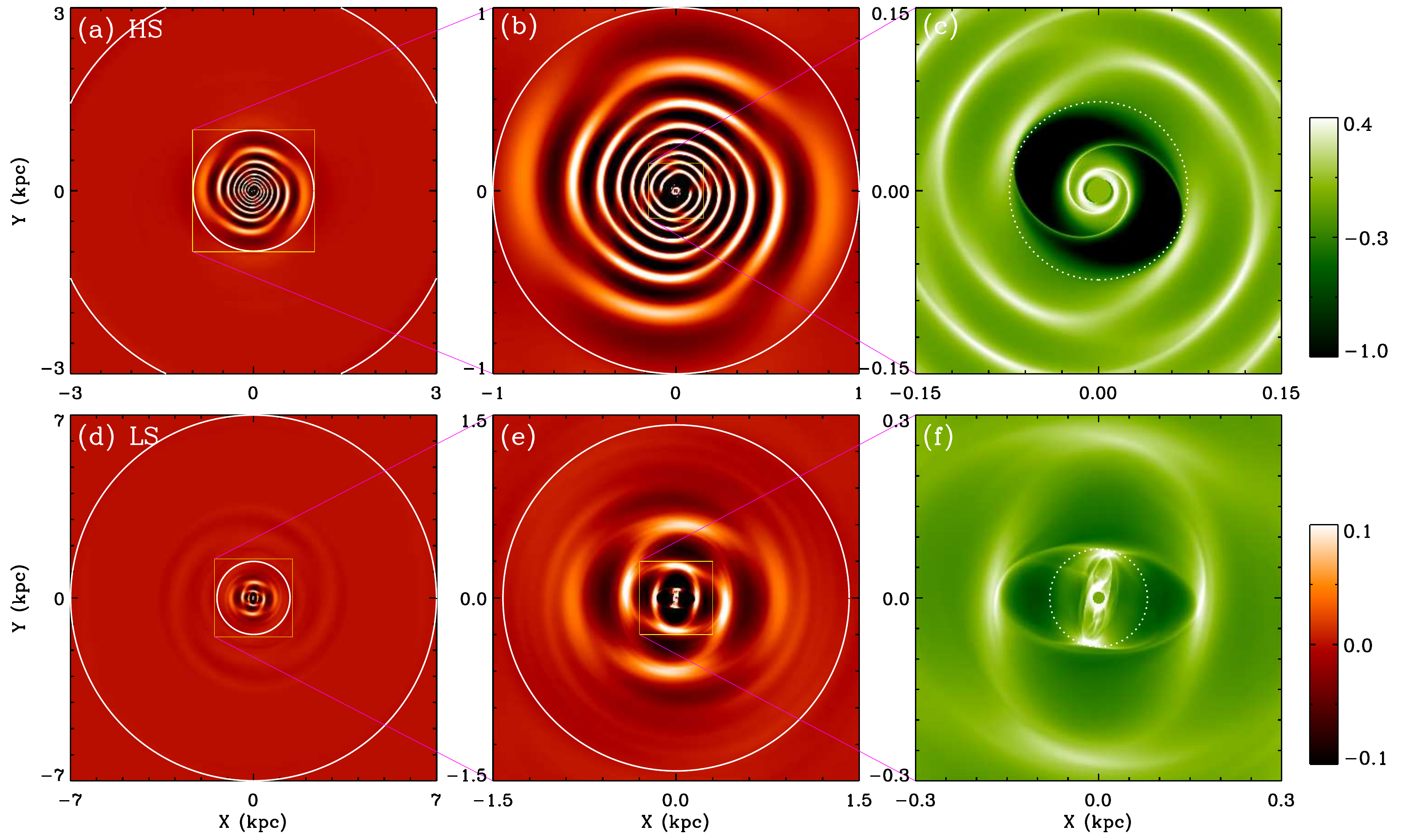}
\caption{Distributions of the gas surface density $\Sigma$ at $t=1$ Gyr for the HS (upper panels) and LS (lower panels) models with $\Phi_a/V_0^2=2\times 10^{-4}$ and $\cs=10\kms$. The two circles in the left panels denote the CO and ILR, with the regions inside the ILR zoomed in to the middle panels. In the middle panels, the central regions bounded by the squares are zoomed in to the right panels. In the right panels, the dotted circle, with radius $R=$ 70 pc in the HS model and 80 pc in the LS model, marks the boundary of the regions influenced by shocks. In the HS model, the inflowing gas is accumulated at small $R$, forming a circumnuclear disk with radius of $\sim30$ pc. The disk is rotating in the counterclockwise direction. Both colorbars label $\log(\Sigma/\Sigma_0)$, with the upper (lower) one mapping for the right (left and middle) panels. }\label{f:snapshot}
\end{figure*}

Figure \ref{f:logRphi} plots the snapshots of the gas surface density $\Sigma$ at selected times for the HS (left) and LS (right) models with $\Phi_a/V_0^2=2\times 10^{-4}$. Figure \ref{f:snapshot} displays the distributions of $\Sigma$ at $t=1.0\Gyr$ in the $x$--$y$ plane, with successive zoom-in views. The nuclear spirals are limited to inside the CR, with very weak density variations outside the CR. They are tightly wound in the HS model due to strong background shear, and loosely wound in the LS model. The spirals are piecewise logarithmic, with a pitch angle of $i_p \sim10^\circ$ at $R> 0.15\kpc$ in the HS model and $i_p\sim 35^\circ$ at $R>0.3\kpc$ in the LS model.

%fig4

Initially, the shape of the nuclear spirals is controlled by the radial gradient of $\Theta\equiv \Omega-\kappa/2$ such that they are leading in the regions where $d\Theta/dR>0$ and trailing where $d\Theta/dR<0$ \citep{buta96,kim12b}, a characteristics of kinematic density waves \citep{wada94,sormani15}. Favored by shear, trailing spirals usually grow faster and more strongly than leading spirals \citep{kim12b}. In addition, the regions (between $R_1$ and $R_2$ in Fig.~\ref{f:logRphi}) for leading spirals are so narrow that the density inside the ILR becomes soon dominated by trailing spirals. Due to a geometric effect \citep{montenegro99}, trailing spirals increase in amplitude as they converge to the origin, analogous to the amplification of sound waves in sonoluminescence \citep{kondic95}. Thus, the nuclear spirals become nonlinear earlier at smaller $R$, readily developing into shock waves.  Strong background shear makes the shock fronts inclined to the gas streamlines in the HS model, resulting in relatively low perpendicular Mach numbers of $\mathcal{M}_\perp\sim 1.5$. In the LS model, the shock fronts are less inclined and become stronger as they unwind over time, resulting in $\mathcal{M}_\perp\sim 2.5$. This unwinding is presumably caused by a nonlinear increase in the angular momentum flux \citep{lee99}.  We find that nuclear spirals in models with $\cs=20\kms$ evolve qualitatively similarly to those in the lower $\cs$ counterparts, although they form earlier in the former due to stronger thermal pressure.

The formation of shocks has two effects: the resulting torques and orbital energy dissipation cause gas accretion to the center, while back reactions to the torques modify the trailing spirals. Because of the first effect, gas encountering the shocks loses its angular momentum and moves inward, eventually passing through the inner radial boundary of our simulation. At the same time, the regions with shocks grow in size radially as the spirals become stronger, reaching $R\sim 70\pc$ at $t=0.5\Gyr$ in the HS model and $R\sim 80\pc$ at $t=0.8\Gyr$ when the spirals saturate. The boundary of the shocked regions at $t=1\Gyr$ is indicated by the dotted circles in Figures \ref{f:snapshot}c and \ref{f:snapshot}f. In the HS model, the spiral shocks are oblique and in a logarithmic shape. The inflowing gas follows more-or-less circular orbits with small-amplitude epicycle motions, as indicated by instantaneous streamlines in Figure \ref{f:logRphi}, and piles up near the center due to the geometric convergence effect, forming a circumnuclear disk with radius of $\sim 20-30\pc$. In the LS model, on the other hand, the shocks are almost perpendicular and gas after the shocks makes quite a radial orbit especially at $R\lesssim25\pc$, plunging almost directly to the inner radial boundary, without forming a circumnuclear disk.

The second effect of the shocks changes the density and velocity fields significantly -- not only in the shocked regions but also the surrounding regions. In particular, the velocity induced by the shocks is able to boost the epicycle amplitudes of gas elements just outside the shocked regions. The perturbations propagate outward in the form of pressure-modified inertial waves and interfere with the trailing spirals there. In the HS model, the perturbations are so weak that their effect is moderate, creating only
mild modulation of the spirals. On the other hand, the perturbations in the LS model are quite strong and make the trailing spirals branch out, resulting in V-shaped structures, similar to those seen in NGC 2207 (Fig.~\ref{f:image}).  After the shocks saturate, gas just outside the shocked regions is pushed inward due to pressure gradients and experiences more shocks successively in orbital motions, moving toward the center. The infall radial velocity is typically $\sim 30\kms$ and $\sim 50\kms$ in the HS and LS models, respectively.

%fig5
\begin{figure*}
\centering\includegraphics[width=6in]{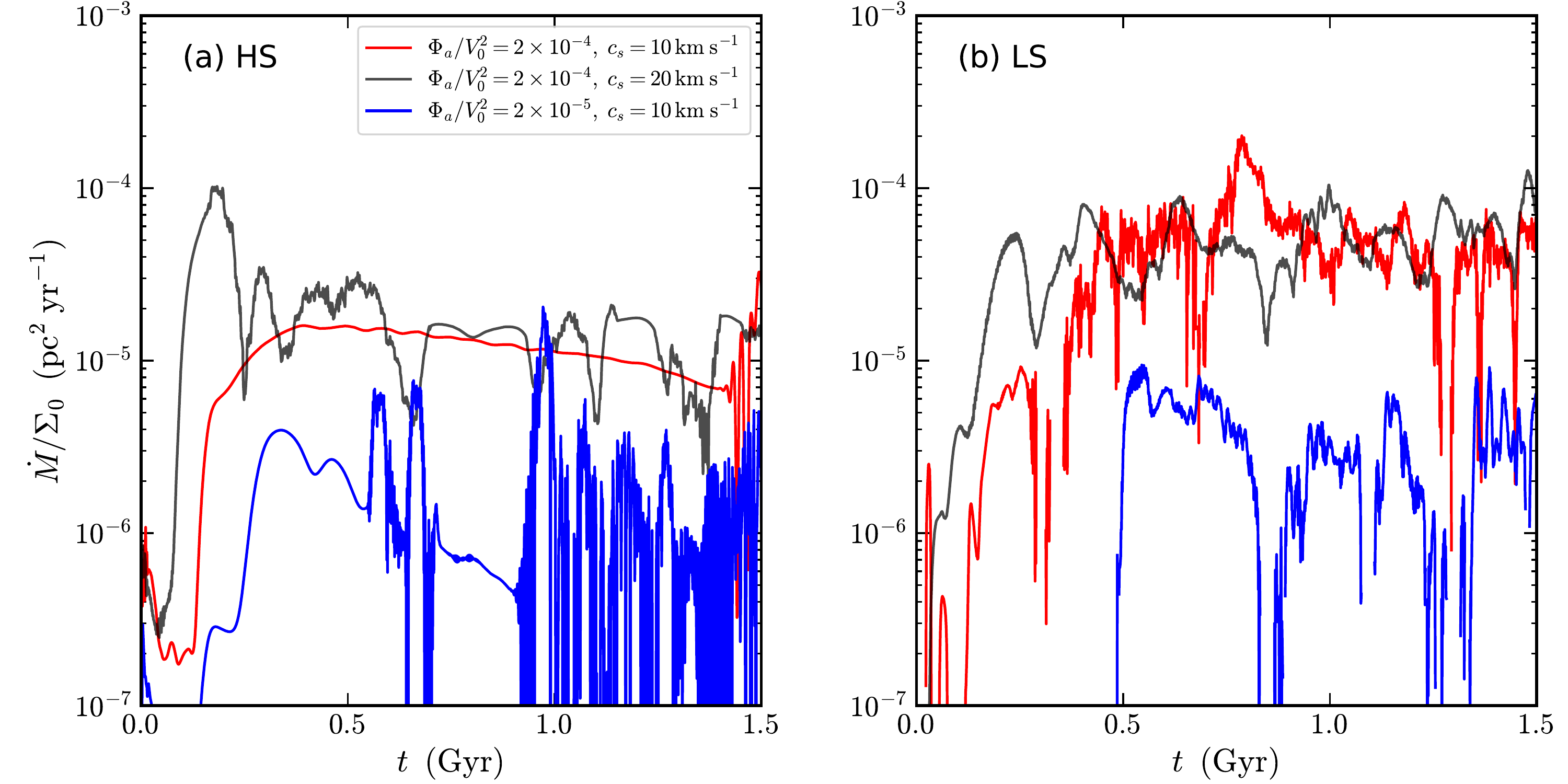}
\caption{Mass inflow rates $\dot{M}$ measured at the inner radial boundary $R_{\rm in}=10\pc$. The red and blue  curves are for the models with $\cs=10\kms$, while the black curves correspond to the models with $\cs=20\kms$. The weak inflows at $t<0.02\Gyr$ are transients caused by a sudden introduction of the external potential. Subsequent massive inflows are primarily driven by shocks. In the HS models, the gas accretion is more or less steady when $\Phi_a/V_0^2=2\times 10^{-4}$, while fluctuating rapidly especially at late time when $\Phi_a/V_0^2=2\times 10^{-5}$. The time-averaged mass inflow rates after $t=0.4\Gyr$ in the HS models are $\langle \dot M\rangle / (\Sigma_0 \,{\rm pc^2\,yr^{-1}}) \sim 10^{-5}$ and $10^{-6}$ for $\Phi_a/V_0^2=2\times 10^{-4}$ and $2\times 10^{-5}$, respectively, insensitive to $\cs$. In the LS models, the mass inflows occur in a  stochastic and intermittent fashion with  $\langle \dot M\rangle / (\Sigma_0 \,{\rm pc^2\,yr^{-1}}) \sim 5 \times 10^{-5}$ and $3 \times 10^{-6}$ for the cases with $\Phi_a/V_0^2=2\times 10^{-4}$ and $2\times 10^{-5}$, respectively, after $t=0.6\Gyr$, again insensitive to $\cs$.}\label{f:mdot}
\end{figure*}

Figure \ref{f:mdot} plots the mass inflow rate $\dot{M} (t) = -\displaystyle\int \Sigma v_R R d\phi$ measured at $R=R_{\rm in}$ divided by $\Sigma_0$, where $v_R$ is the radial velocity, for all models with both $\cs=10$ and $20\kms$. In the HS models, the nuclear spirals grow fast and initiate gas inflows when the shocks first form near the inner boundary. Since the spirals are tightly wound, however, the shocks are relatively weak and the induced mass inflow rate remains small in these models. In the LS models, on the other hand, spirals are more loosely wound and grow more slowly, resulting in stronger shocks and larger gas inflows. Note that the mass inflow rate in our models with a weak bar/oval is insensitive to $\cs$. This is in sharp contrast to the cases with a strong bar in which thermal pressure spreads out the gas in a nuclear ring to enhance the gas density at the center, resulting in larger $\dot{M}$ for larger $\cs$ \citep{ann05,kim12b}. In all models, the mass inflow rate is almost proportional to the amplitude of the imposed external potential, and persists for a long period of time.

The time-averaged mass inflow rates in our models can be summarized as $\langle \dot M\rangle = 10^{-5} f\Sigma_0 (\Phi_a V_0^{-2}/2\times 10^{-4})$ $M_\odot$ yr$^{-1}$, where $f$ is a factor, varying less than an order of magnitude, arising from the difference in the background shear, and $\Sigma_0$ is in $\surf$. If all the inflowing mass is accreted to an SMBH with mass $M_{\rm BH}$, this corresponds to an Eddington ratio of
\begin{equation}
 \begin{split}
 \lambda = \frac{L_{\rm bol}}{L_{\rm Edd}} =
  1.5\times10^{-2}f
  \left(\frac{\epsilon}{0.1}      \right)
  \left(\frac{\Phi_a V_0^{-2}}{2\times 10^{-4}}\right)\qquad \\
  \left(\frac{\Sigma_0} {10^2\surf}\right)
  \left(\frac{M_{\rm BH}}{3\times10^6\Msun}\right)^{-1},
  \end{split}
\end{equation}
where $L_{\rm bol}=\epsilon \langle \dot M\rangle c^2$ is the bolometric luminosity of an AGN with efficiency $\epsilon$, and $L_{\rm Edd}$ is the Eddington luminosity, $3.2\times10^4\,(M_{\rm BH}/M_\odot)\,L_\odot$. Observations show that $\lambda \lesssim 0.1$ for classical Seyfert 1 galaxies with broad emission lines \citep{meyer11}, $\lambda \sim 10^{-2}$ for Seyfert 2 galaxies with narrow lines \citep{bian07}, and $\lambda \sim 10^{-3} $ for low-luminosity Seyfert 1 galaxies \citep{ho08}. These values suggest that gas inflows driven by shock fronts in nonlinear nuclear spirals can account for observed levels of AGN activity in Seyfert galaxies, depending on the
amount of gas in the disk and the amplitude of the external potential.

\section{Turbulence Generation}

Accretion at a rate ${\dot M}$ corresponds to a rate of energy input to the gas equal to about $\case{1}{2}{\dot M} V^2$ for orbital speed $V$ \citep[see][for a more detailed description]{krumholz15}. The energy dissipation rate from a uniform disk with surface density $\Sigma$, radius $R$, mass $M=\pi R^2 \Sigma$, velocity dispersion $\sigma$, and scale height $H=\sigma^2/(\pi G \Sigma)$ is assumed to equal a factor $\delta <1$ times the turbulent energy divided by the crossing time over a disk thickness: $\case{1}{2} \delta M \sigma^3/H = \case{1}{2} G \pi^2 \delta R^2 \Sigma^2 \sigma$. If we set the energy input rate equal to the energy dissipation rate and rearrange, we get the velocity dispersion in the inner disk,
\begin{equation}
  \sigma \sim \frac {({\dot M}/\Sigma)V^2} {G \pi^2 \delta R^2 \Sigma}\,.
\end{equation}
Figure \ref{f:mdot} shows that ${\dot M}/\Sigma \sim10^{-5}\pc^{2}\yr^{-1}$ for $\Phi_a/V_0^{2}=2\times 10^{-4}$. Inserting standard values, we obtain
\begin{equation}
  \sigma\sim 230 \frac{V_{100}^2}{\delta R_{10}^2 \Sigma_{100}} \,\kms,
\end{equation}
where $V_{100}=V/(100\kms)$, $R_{10}=R/(10\pc)$, and $\Sigma_{100}=\Sigma/(100\surf)$.  For radii up to tens of parsecs, $\sigma$ exceeds $\sim20\kms$. For realistic $\delta<1$, $\sigma$ is even higher.

This high value of $\sigma$ proposed for the three-dimensional case suggests that the accretion rate in our 2D simulations is high enough to pump supersonic turbulence at small radii. Such turbulence would presumably give a Kolmogorov power spectrum like that observed for nuclear dust structure in NGC 4450 and NGC 4736 \citep{elmegreen02}. Our simulations do not show it directly because the dissipation rate in the simulations is much higher than in the derivation above, as the model gas is forced to retain the same velocity dispersion (i.e., isothermal) with disk thickness $H=0$.

\section{Discussion}

We have presented the results of 2D hydrodynamic simulations for nuclear spirals driven by an oval distortion or a weak bar potential. Due to the geometric effect, nuclear spirals turn to shocks at small radii even if the potential perturbations are very weak. The background shear affects the morphologies of the nuclear spirals and shocks significantly, such that spirals are tightly (loosely) wound and the shocks are oblique (perpendicular) when shear is high (low). Our high-shear HS model forms a circumnuclear disk embedded with loosely-wound spiral shocks by accumulating inflowing gas. Circumnuclear disks with spirals seen in external galaxies, for example, in NGC 1097 \citep{hsieh11,onish15}, could form by a similar process. We note however that NGC 1097 and Milky Way are strongly barred galaxies, so that nuclear rings produced by the strong bars might have affected the formation of circumnuclear disks.

Most previous theoretical work that studied gas accretion onto SMBHs considered giant elliptical galaxies embedded in the hot gaseous halos of their galaxy clusters, finding that gas accretion is chaotic and intermittent, occurring as the cold clouds formed by cooling and thermal instability in the hot gas rain down \citep{pizzolato05,sharma12,gaspari13,voit15}. Indeed, a cold, clumpy accretion flow has recently been observed in the nucleus of the brightest cluster galaxy in Abell 2597 \citep{tremblay16}. Our results suggest that nuclear spiral shocks at the centers of gas-rich disk galaxies, rather than thermal processes, can feed their SMBHs. These shocks are limited to the very central regions where driven pressure waves converge and steepen to become nonlinear. The radial inflow velocity of $\sim50\kms$ observed in the center of NGC 1097 \citep{fathi06} is consistent with our model.

\acknowledgements
We are grateful to D.~Elmegreen for help in making Figure 1 and to J.~Stone for permission to use the Athea++ code at the developing stage of the code. This work was supported by the National Research Foundation of Korea
(NRF) grant funded by the Korean government (MEST) (No. 3348-20160021). The computation of this work was performed on the Linux cluster at KASI (Korea Astronomy and Space Science Institute).

\end{document}